\documentclass[a4paper,11pt]{article}
\pdfoutput=1

\usepackage{jheppub}
\usepackage{booktabs}
\usepackage{xcolor}
\usepackage{bm}

\newcommand{\bea} {\begin{eqnarray}}
\newcommand{\eea} {\end{eqnarray}}

\newcommand{\beq} {\begin{equation}}
\newcommand{\eeq} {\end{equation}}

\newcommand{\order}{{\cal O}}

\def\hm{{\hat m}}
\makeatletter
\gdef\@fpheader{}
\makeatother

\title{Electroweak phase transition triggered by a long-lived scalar and LHC searches}

\author[a]{Lei Wang,}
\author[a]{Xiaoting Zhao}
\author[a,1]{Xiao-Fang Han\note{Corresponding author.}}
\author[a,2]{Hongxin Wang\note{Corresponding author.}}

\affiliation[a]{Department of Physics, Yantai University, Yantai 264005, China}

\emailAdd{leiwang@ytu.edu.cn}
\emailAdd{zhaoo\_xiaoting@126.com}
\emailAdd{xfhan@ytu.edu.cn}
\emailAdd{hxwang@ytu.edu.cn}

\abstract{
We investigate a long-lived scalar $S$ that triggers a strongly first-order electroweak phase transition (SFOEWPT) in the singlet scalar extension of standard model (SM). A sufficient small mixing angle between $S$ and the SM-like Higgs boson $h$ naturally leads to a long lifetime for $S$.
We focus on the mass range $m_S>m_h/2$, where the dominant production channel at the LHC is the Higgs-mediated process $gg\to h^*\to SS$ with an off-shell $h$.
 By analyzing the finite-temperature effective potential, we identify the parameter space realizing an SFOEWPT and calculate the associated gravitational-wave spectra, finding that part of the parameter space can be probed by LISA. We further study the LHC sensitivity to the long-lived $S$ using the ATLAS detector and proposed far detectors, including ANUBIS, CODEX-b, FACET, FASER2, MoEDAL-MAPP1 and MAPP2, and MATHUSLA.
The combined sensitivity of the main detector and far detectors, 
particularly ANUBIS and MATHUSLA, 
allows a large fraction of the SFOEWPT parameter space to be probed for $10^{-10}\leq \sin\theta\leq 10^{-5}$, with $m_S$ up to $250~\mathrm{GeV}$. 
For $\sin\theta\leq 10^{-11}$, however, the SFOEWPT parameter space becomes inaccessible to both the main detector and the far detector searches.
 }

\begin{document} 
\maketitle
\flushbottom

\section{Introduction}
\label{sec:intro}

The electroweak phase transition (EWPT) provides an important connection between particle physics and the early Universe. It plays a crucial role in electroweak baryogenesis \cite{Kuzmin:1985mm,Rubakov:1996vz}, which requires a strongly first-order electroweak phase transition (SFOEWPT) to generate the observed baryon asymmetry through the departure from thermal equilibrium. However, the standard model (SM) with the observed Higgs boson mass of approximately $125$ GeV predicts a crossover transition rather than a first-order phase transition. Therefore, new physics beyond the SM is required to realize an SFOEWPT. 
Among various extensions of the SM, an enlarged scalar sector provides a simple framework. 
The additional scalar degrees of freedom can modify the  effective scalar potential and lead to distinctive collider signatures.
In many studies, the additional scalars are assumed to decay promptly through their couplings to the SM particles. Such prompt-decaying scalars can be searched for at the Large Hadron Collider (LHC) through various channels \cite{Carena:2019une,Bernon:2017jgv,Wang:2018hnw,Wang:2019pet,Carena:2022yvx,Goncalves:2022wbp,Bosse:2026bdk,Agrawal:2023cgp,Blasi:2022woz,Sengupta:2025jah}. To date, the ATLAS and CMS collaborations have reported no significant excess pointing to new scalar states, thereby imposing stringent constraints on the parameter space of scalar extensions.

A different and less explored possibility arises when the scalar particle has a sufficiently small coupling to the SM particles. In this case, the scalar decay width can be highly suppressed, resulting in a macroscopic lifetime. A long-lived particle (LLP) can simultaneously affect the thermal evolution of the early Universe and produce experimentally accessible displaced signatures at the collider. This establishes an interesting connection between the origin of the EWPT and LLP searches at high-energy colliders. Besides ATLAS and CMS, far detectors such as ANUBIS \cite{Bauer:2019vqk,ANUBIS:2025sgg}, CODEX-b \cite{Gligorov:2017nwh,CODEX-b:2019jve}, FACET \cite{Cerci:2021nlb}, FASER \cite{Feng:2017uoz,FASER:2018eoc,FASER:2022hcn}, FASER2 \cite{FASER2}, MoEDAL-MAPP1 and MAPP2 \cite{Pinfold:2019nqj,Pinfold:2019zwp}, and MATHUSLA \cite{Curtin:2018mvb,Chou:2016lxi,MATHUSLA:2018bqv,MATHUSLA:2020uve,MATHUSLA:2025zyt}, can extend the sensitivity to LLPs with macroscopic decay lengths.

A widely studied production mechanism for long-lived scalar ($S$) is through the decay of an on-shell SM-like Higgs boson ($h$), 
$h\to SS$ (see, e.g., Refs.~\cite{Alipour-Fard:2018lsf,Filimonova:2019tuy,Cheung:2019qdr,Liu:2022nvk,Wang:2024ieo,Cepeda:2021rql,Guo:2025qes}). 
Such signatures have been extensively searched for by the ATLAS and CMS collaborations \cite{ATLAS:2022gbw,ATLAS:2025pak,CMS:2024bvl}, which have constrained the corresponding parameter space.
However, the searches based on on-shell Higgs decays are limited to the mass range below $m_h/2$. In Ref. \cite{Wang:2026ifw}, the authors first proposed 
the off-shell Higgs-mediated production mechanism, $gg\to h^*\to SS$, which explores heavier long-lived scalars beyond the Higgs decay threshold.

In this work, we explore a long-lived  $S$ that triggers an SFOEWPT in the real singlet scalar extension of SM. 
The scalar $S$ mixes with the SM-like Higgs boson $h$, and a sufficient small mixing angle naturally leads to a long lifetime for $S$. 
We focus on the scalar mass range above $m_h/2$, where the dominant production mechanism at the LHC is the off-shell Higgs-mediated process ($gg \to h^* \to SS$). The complementary scenario in which $S$ is produced through the decay of an on-shell $h$ has been studied as a probe of the SFOEWPT in \cite{Liu:2022nvk}.
 We analyze the SFOEWPT and calculate the corresponding gravitational wave (GW) spectra. 
Furthermore, we investigate detection prospects of the long-lived $S$ at the ATLAS detector and several proposed far detectors, including  ANUBIS, CODEX-b, FACET,  FASER2, MoEDAL-MAPP1 and MAPP2, and MATHUSLA.

The paper is organized as follows. In Sec.~\ref{sec:model}, we briefly introduce the model. 
Sec.~\ref{sec:decay} is devoted to the decay properties of the long-lived $S$ and the relevant constraints. 
In Sec.~\ref{sec:EWPT}, we study the SFOEWPT and the corresponding spectra. 
 The detection prospects of the long-lived $S$ at the ATLAS detector and several far detectors
 are presented in Sec.~\ref{sec:lhc}. Finally, we summarize our conclusions in Sec.~\ref{sec:conclu}.

\section{The real singlet scalar extension of the standard model}
\label{sec:model}

The SM is extended by introducing a real singlet scalar field $S$. Without imposing a $Z_2$ symmetry, the most general renormalizable scalar potential is given by
\begin{eqnarray} \label{VSMS} \mathrm{V} =&&-\mu_h^2
(\Phi^{\dagger} \Phi) + \frac{\lambda_1}{2}  (\Phi^{\dagger} \Phi)^2 +
\frac{\mu_2}{2} \mathbb{S}^2 + \frac{\lambda_2}{8}  \mathbb{S}^4 + \mu_1 \mathbb{S} +
\frac{\mu_3}{3} \mathbb{S}^3 \nonumber\\
&&+ \frac{\kappa_1}{2} \Phi^{\dagger} \Phi \mathbb{S} +  \frac{\kappa_2}{2} \Phi^{\dagger} \Phi \mathbb{S}^2,
\end{eqnarray}
where $\Phi$ denotes the Higgs doublet field and $\mathbb{S}$ is the real singlet scalar field. 
We set $\mu_3=0$ by a shift invariance of the singlet potential.

After electroweak symmetry breaking, the fields are expanded around the vacuum as
\begin{equation}
\Phi=
\begin{pmatrix}
G^+\\
\frac{1}{\sqrt{2}}(v+\phi^0+iG^0)
\end{pmatrix},
\qquad
\mathbb{S}=s,
\label{eq:field_expansion}
\end{equation}
where $v\simeq246$ GeV is the Higgs vacuum expectation value (VEV), and we take the singlet VEV to be zero, $\langle \mathbb{S}\rangle=0$.
The conditions for minimizing the scalar potential lead to,
\begin{equation}
\mu_h^2=\frac{\lambda_1}{2} v^2,~~\mu_1=-\frac{\kappa_1}{4} v^2.
\label{eq:min_H}
\end{equation}

The physical scalar states arise from the mixing of $\phi^0$ and $s$,
\begin{eqnarray}
\left(\begin{array}{c}h \\ S \end{array}\right) =  \left(\begin{array}{cc}\cos\theta & \sin\theta \\ -\sin\theta & \cos\theta \end{array}\right)  \left(\begin{array}{c} \phi^0 \\ s \end{array}\right),
\end{eqnarray}
where the shorthand notations are $c_\theta \equiv \cos\theta$ and $s_\theta\equiv \sin\theta$. 
The $h$ is is identified as the observed SM-like Higgs boson with $m_h=125$ GeV, while $S$
denotes the additional scalar state.
Through its mixing with $h$, the scalar $S$ acquires couplings to the SM fermions and gauge bosons as
\begin{align}
\mathcal{L}_{S} = \frac{m_f}{v}\sin\theta \bar{f}fS-2\frac{m^2_W}{v}\sin\theta W_{\mu}^+W^{-\mu} S -\frac{m^2_Z}{v}\sin\theta Z_{\mu}Z^{\mu} S.
\label{effcoup}
\end{align}

The parameters $\lambda_1$, $\kappa_1$, and $\mu_2$ are expressed in terms of the scalar masses and mixing angle,
\begin{eqnarray}
&& \lambda_1  = \frac{ m_{h}^2 c_\theta^2+m_{S}^2 s_{\theta}^2}{v^2},  ~~~\kappa_1 = \frac{ 2 (m_{h}^2 - m_S^2) s_\theta c_\theta}{v},  \nonumber \\  
&&\mu_2 =  -\frac{\kappa_2}{2}v^2 + m_h^2 s_\theta^2 +m_S^2 c_\theta^2. 
\label{eq:lambdas}
\end{eqnarray}

The cubic scalar interactions between $h$ and $S$ contains
\beq\label{lag-cubic}
\mathcal{L}_{\text{cubic}} = - \frac{\lambda_{hSS}}{2} v hSS - \frac{\lambda_{Shh}}{2} v Shh,
\eeq
with
\beq
\lambda_{hSS} = \kappa_2  c_\theta^3, ~~~
\lambda_{Shh} = 2 \kappa_2  c_\theta^2 s_\theta - \frac{ 2m_h^2   + m_S^2 }{v^2} c_\theta^4 s_\theta.
\eeq
Here, we retain terms only up to $\order(s_\theta)$. For a sufficiently small $s_\theta$, Eqs. (\ref{effcoup}) and (\ref{lag-cubic}) show that the decay width of $S$ is highly suppressed, resulting in a long lifetime. Meanwhile, from Eqs. (\ref{eq:min_H}) and (\ref{eq:lambdas}), both $\mu_1$ and $\kappa_1$ are proportional to $s_\theta$. Since these terms break the $Z_2$ symmetry, a small $s_\theta$ is naturally protected by an approximate $Z_2$ symmetry.

\section{Long-lived scalar and relevant constraints}
\label{sec:decay}
In this section, we first discuss the theoretical constraints on the parameter space of the model, including vacuum stability and perturbative unitarity.
The stability of the scalar potential at large field values is determined by its quartic terms. 
For the potential to be bounded from below, the quartic part must remain positive for arbitrary values of $\Phi$ and $\mathbb{S}$. This leads to the conditions \cite{Kanemura:2015fra,He:2016sqr}
\begin{equation}
\lambda_1>0,
\qquad
\lambda_2>0,
\qquad
\kappa_2>-\sqrt{\lambda_1\lambda_2}.
\label{eq:vacuum_stability}
\end{equation}
In addition to the bounded-from-below conditions, we require the electroweak vacuum to be the global minimum of the scalar potential. 
Since the exact analytical condition for the global minimum is difficult to obtain, we impose this condition numerically in our analysis.

\begin{figure}[tb]
\begin{center}
 \epsfig{file=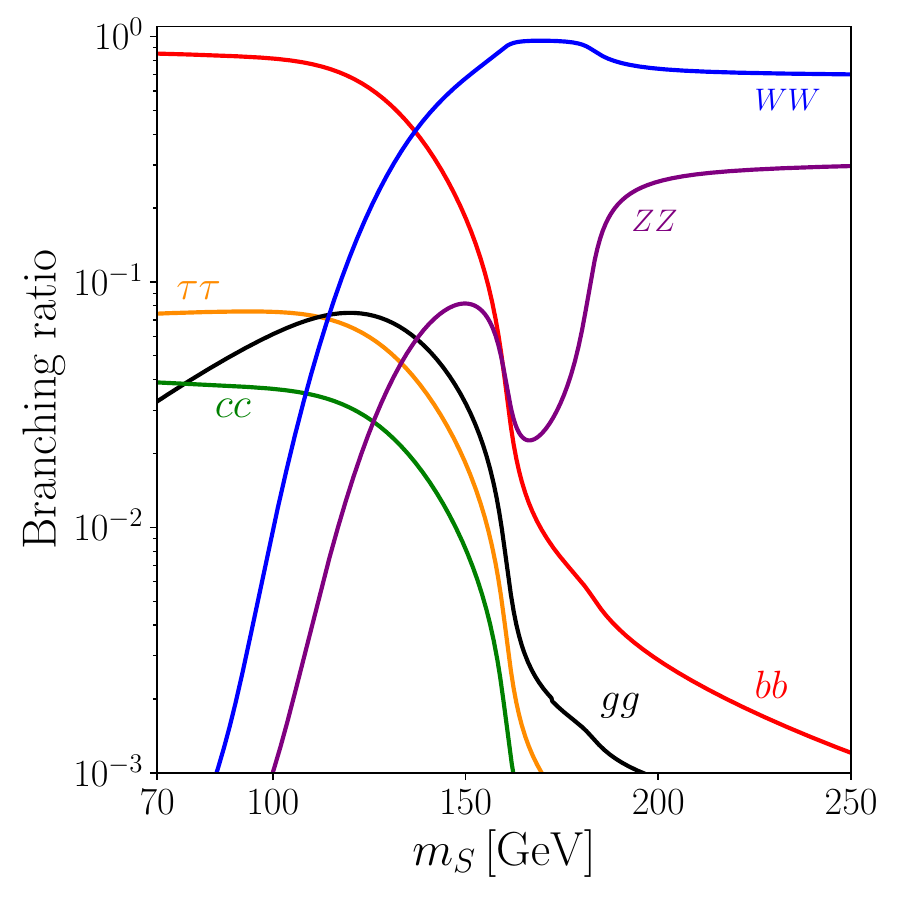,height=7.5cm}
 \end{center}
\vspace{-1.0cm} \caption{The branching ratio of the different $S$ decay modes.} \label{fig:br}
\end{figure}

\begin{figure}[tb]
\begin{center}
\epsfig{file=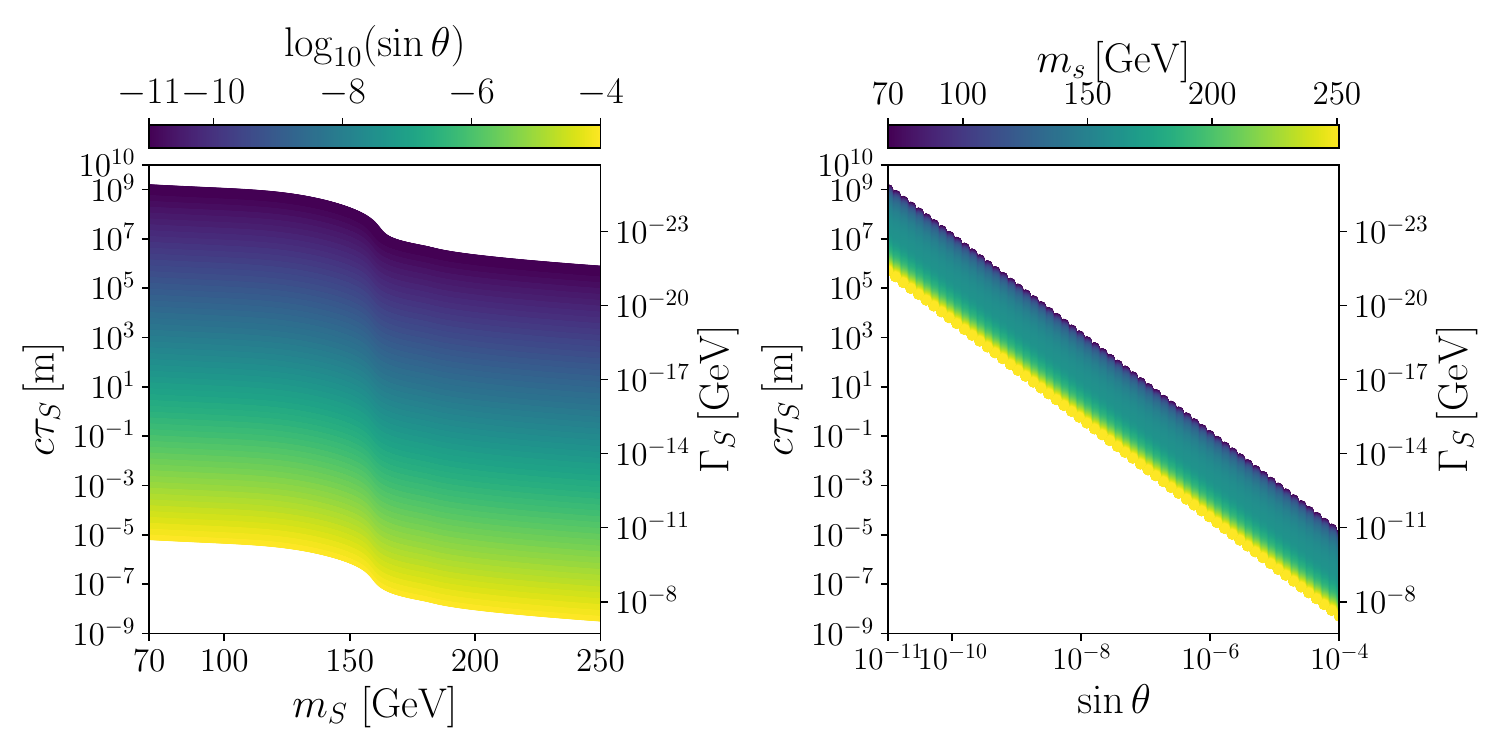,height=7.5cm}
 \end{center}
\vspace{-1.0cm} \caption{The proper decay length $c\tau_S$ as a function of $m_S$ and $\sin\theta$, with the total decay width of $S$ displayed on the right $y$-axis.} \label{fig:wid}
\end{figure}

We further impose perturbative unitarity constraints on the scalar quartic couplings. 
At high energies, the scattering amplitudes of scalar fields are dominated by the quartic interactions,
which lead to the following conditions  \cite{Kanemura:2015fra,He:2016sqr},
\bea
&&\lambda_1 < 8\pi,
\qquad
\kappa_2 < 8\pi,\nonumber\\
&&
\frac{1}{2}\left|\frac{3}{2}\lambda_2+3\lambda_1 \pm \sqrt{4\kappa_2^2 +\frac{9}{4}\lambda_2^2+9\lambda_1^2-9\lambda_2\lambda_1}\right| < 8\pi.
\label{eq:vacuum_stability}
\eea

The partial decay width of $S\to f\bar{f}$, $WW$, $ZZ$, $\gamma\gamma$ and $gg$ is calculated as 
\beq
\Gamma_{h\to XX}= s^2_\theta \times \Gamma_{h\to XX}^{SM},
\label{eqdecay}
\eeq
where $\Gamma_{h\to XX}^{\rm SM}$ is the partial decay width evaluated with the SM couplings.
Once $m_S>250$ GeV, the $S\to hh$ channel opens, and it can  affect the lifetime of 
$S$, despite the suppression of the $Shh$ coupling by $s_\theta$. 
Including this channel would complicate the lifetime analysis, we therefore restrict our study to the mass range $m_S\leq250$ GeV for simplicity.

For a sufficiently small $s_\theta$, the scalar $S$ can have a macroscopic lifetime and behave as an LLP. 
Since all the partial decay widths in Eq. (\ref{eqdecay}) share the same $s_\theta^2$ dependence, 
the corresponding branching ratios are insensitive to the value of $s_\theta$.
We modify the \textbf{2HDMC} \cite{Eriksson:2009ws} package to evaluate the decay widths of $S$, and display
the branching fractions in the Fig. \ref{fig:br}.
With increasing $m_S$, the dominant decay mode shifts from $S\to b\bar{b}$ to $S\to WW$.
Also, the branching fraction of $S\to ZZ$ can exceed that of $S\to b\bar{b}$,
 mainly as a consequence of the opening of the gauge-boson decay channels 
and the corresponding enhancement of the gauge-boson couplings.

Fig.~\ref{fig:wid} shows the proper decay length $c\tau_S$ as a function of $m_S$ and $s_\theta$. The total decay width of $S$ scales as $s_\theta^2$ and decreases rapidly as $s_\theta$ becomes smaller. Meanwhile, the decay width increases with $m_S$, particularly after the opening of the $S\to W^+W^-$ channel, followed by the $S\to ZZ$ channel at higher masses. Consequently, the decay length $c\tau_S$ increases as both $m_S$ and $s_\theta$ decrease, reaching the macroscopic scale and exceeding one meter for sufficiently small $m_S$ or $s_\theta$.


\section{Electroweak phase transition}
\label{sec:EWPT}
\subsection{Finite-temperature effective potential and bubble nucleation}
The additional scalar field can significantly modify the effective scalar potential and consequently affect the thermal history of the Universe. 
The neutral elements of $\Phi$ is parametrized as $\frac{\phi}{\sqrt{2}}$, and the singlet scalar field is parametrized as $s$.
The finite-temperature effective potential at the one-loop level can be written as
\begin{align}
V_{\rm eff}(\phi,s,T)
={}&V_0(\phi,s)
+V_{\rm CW}(\phi,s)
+V_{\rm CT}(\phi,s)
\nonumber\\
&+V_T(\phi,s,T)
+V_{\rm ring}(\phi,s,T),
\label{veff0}
\end{align}
where $V_0$, $V_{\rm CW}$, $V_{\rm CT}$, $V_T$, and $V_{\rm ring}$ denote the tree-level contribution, the Coleman-Weinberg correction \cite{Coleman:1973jx}, the counterterm contribution, the finite-temperature correction \cite{Dolan:1973qd}, and the daisy resummation contribution \cite{Arnold:1992rz,Parwani:1991gq}, respectively. We evaluate the effective potential in the Landau gauge.

In terms of the classical background fields $(\phi,s)$, the tree-level potential takes the form
\begin{eqnarray}
V_{\rm 0}(\phi,s)&=&
-\frac{1}{2} \mu_h^2 \phi^2 + \mu_1 s + \frac{1}{2} \mu_2 s^2 + \frac{1}{3} \mu_3 s^3
\nonumber\\
&&+\frac{1}{8} \lambda_1 \phi^4
+\frac{1}{8} \lambda_2 s^4
+\frac{1}{4}\kappa_1 \phi^2  s
+\frac{1}{4}\kappa_2 \phi^2 s^2.
\end{eqnarray}

The one-loop Coleman-Weinberg contribution in the $\overline{\rm MS}$ scheme is given by \cite{Coleman:1973jx},
\begin{equation}
V_{\rm CW}(\phi,s)
=
\sum_i
(-1)^{2s_i}
n_i
\frac{\hm_i^4(\phi,s)}{64\pi^2}
\left[
\ln\frac{\hm_i^2(\phi,s)}{Q^2}
-C_i
\right],
\label{eq:CWpot}
\end{equation}
where the sum runs over $i=h,S,G,G^\pm,W^\pm,Z,t$, with $s_i$ denoting the spin of the particle $i$. Here, $\hm_i(\phi,s)$ represents the corresponding field-dependent mass, which is listed in Appendix~A. We choose the renormalization scale as $Q^2=v^2$. The constants $C_i$ are given by $C_i=3/2$ for scalars and fermions, while $C_i=5/6$ for gauge bosons. The corresponding numbers of degrees of freedom are
\begin{align}
&n_h=n_S=n_G=1,\qquad n_{G^\pm}=2,\nonumber\\
&n_{W^\pm}=6,\qquad n_Z=3,\qquad n_t=12.
\end{align}

The inclusion of $V_{\rm CW}$ modifies the zero-temperature minimization conditions and the CP-even scalar mass matrix. We therefore introduce counterterm to preserve the tree-level vacuum and scalar mass spectrum at $T=0$,
\begin{align}
V_{\rm CT}(\phi,s)
={}&
\delta\mu_h^2\phi^2
+\delta\mu_2s^2
+\delta\lambda_1\phi^4
+\delta\lambda_2s^4
+\delta\mu_1 s
\nonumber\\
&+\delta\kappa_1\phi^2s
+\delta\kappa_2\phi^2s^2.
\label{eq:Vct}
\end{align}
The counterterm coefficients are fixed by requiring that both the first and second derivatives of the one-loop corrected potential retain their tree-level values at the electroweak vacuum,
\begin{align}
\frac{\partial V_{\rm CT}}{\partial\phi}
=
-\frac{\partial V_{\rm CW}}{\partial\phi},&
~
\frac{\partial V_{\rm CT}}{\partial s}
=
-\frac{\partial V_{\rm CW}}{\partial s},\nonumber\\
\frac{\partial^2V_{\rm CT}}{\partial\phi^2}
=
-\frac{\partial^2V_{\rm CW}}{\partial\phi^2},~
\frac{\partial^2V_{\rm CT}}{\partial s^2}
&=
-\frac{\partial^2V_{\rm CW}}{\partial s^2},~
\frac{\partial^2V_{\rm CT}}{\partial\phi\partial s}
=
-\frac{\partial^2V_{\rm CW}}{\partial\phi\partial s}.
\label{eq:V2der}
\end{align}
All these conditions are evaluated at the electroweak vacuum, $\phi=v$ and $s=0$. 
There are seven counterterm coefficients in Eq.~(\ref{eq:Vct}), while only five independent conditions are imposed by Eq.~(\ref{eq:V2der}). We therefore set
$\delta\kappa_2=0$ and $\delta\lambda_2=0$,
and determine the remaining five coefficients,
$\delta\mu_h^2$, $\delta\mu_2$, $\delta\lambda_1$, $\delta\mu_1$, and $\delta\kappa_1$, from the above conditions.

A further complication arises from the infrared behavior of the Coleman-Weinberg potential. 
At $T=0$, the second derivatives of $V_{\rm CW}$ contain logarithmic divergences associated with the vanishing masses of the Goldstone bosons. 
We regulate these infrared divergences by introducing an effective  cutoff,
$m_{\rm IR}^2=m_h^2$,
for the Goldstone masses appearing in the divergent terms. 
This prescription provides a good approximation to the full on-shell renormalization procedure, as discussed in Ref.~\cite{Cline:2011mm}.

The finite-temperature contribution to the effective potential is given by 
\beq
\label{potVth}
 V_{\rm th}(\phi,s,T) = \frac{T^4}{2\pi^2}\, \sum_i n_i J_{B,F}\left( \frac{ \hm_i^2(\phi,s)}{T^2}\right)\;,
\eeq
where $i=h,S,G,G^\pm,W^\pm,Z,t$, and the functions $J_{B,F}$ are 
\beq
\label{eq:jfunc}
J_{B,F}(y) = \pm \int_0^\infty\, dx\, x^2\, \ln\left[1\mp {\rm exp}\left(-\sqrt{x^2+y}\right)\right].
\eeq

To account for the infrared-sensitive bosonic modes, we further include the daisy resummation contributions,
\beq
V_{\rm ring}\left(\phi,s, T\right) =-\frac{T}{12\pi }\sum_{i} n_{i}\left[ \left( \bar{M}_{i}^{2}\left(\phi,s,T\right) \right)^{\frac{3}{2}}-\left( \hm_{i}^{2}\left(\phi,s,T\right) \right)^{\frac{3}{2}}\right] ,
\label{eq:daisy}
\eeq
where the sum includes $i=h,S,G,G^\pm,W^\pm_L,Z_L,\gamma_L$. Here, $W^\pm_L$, $Z_L$, and $\gamma_L$ denote the longitudinal components of the gauge bosons, with
$n_{W^\pm_L}=2$ and $n_{Z_L}=n_{\gamma_L}=1$.
The corresponding thermal Debye masses $\bar{M}_i^2(\phi,s,T)$ are listed in Appendix~A.

For a first-order phase transition, bubbles of the true vacuum nucleate through thermal tunneling. The nucleation rate per unit volume at finite temperature can be expressed as \cite{Affleck:1980ac,Linde:1981zj}
\begin{equation}
\Gamma(T)\simeq A(T)e^{-S_3(T)/T},
\end{equation}
where $A(T)\sim T^4$ denotes the prefactor and $S_3(T)$ is the three-dimensional Euclidean action,
%
	%
		%
		%
		%
	%
	%
The nucleation temperature $T_n$ is determined by requiring an order-one probability for bubble formation within a Hubble volume. This condition can be approximated by
$\frac{S_3(T)}{T}|_{T = T_n}\approx 140$.
We identify an SFOEWPT by imposing
\begin{equation}
\frac{\langle\phi\rangle_n}{T_n} \geq 1,
\label{eq:SFOEWPT}
\end{equation}
where $\langle\phi\rangle_n$ is the Higgs VEV evaluated at the nucleation temperature. We employ \textsf{CosmoTransitions} \cite{Wainwright:2011kj} to analyze the EWPT and obtain the relevant phase-transition quantities.



\subsection{Gravitational wave}
A first-order phase transition can generate GWs through three mechanisms, 
including bubble wall collisions, sound waves in the plasma, and magnetohydrodynamic (MHD) turbulence. The resulting GW spectrum can be written approximately as
\begin{equation}
\Omega_{\text{GW}}h^{2}
\simeq
\Omega_{\rm col}h^{2}
+
\Omega_{\rm sw}h^{2}
+
\Omega_{\rm turb}h^{2}.
\end{equation}
Only a small fraction of the released energy is transferred to the bubble walls
despite the possibility that bubble walls may become relativistic and even undergo runaway expansion in certain scenarios 
 \cite{Bodeker:2009qy}. 
As a result, the GW contribution from bubble wall collisions is generally subdominant and can be neglected \cite{Bodeker:2017cim}. 
We therefore omit $\Omega_{\rm col}$ in our analysis and focus on the contributions from sound waves and MHD turbulence.

The GW spectrum induced by sound waves is given by \cite{Hindmarsh:2015qta}
\begin{eqnarray}
\Omega_{\textrm{sw}}h^{2} & \ = \ &
2.65\times10^{-6}\left( \frac{H_{n}}{\beta}\right)\left(\frac{\kappa_{v} \alpha}{1+\alpha} \right)^{2}
\left( \frac{100}{g_{\ast}}\right)^{1/3} \tilde{v}_W\nonumber \\
&&\times  \left(\frac{f}{f_{sw}} \right)^{3} \left( \frac{7}{4+3(f/f_{\textrm{sw}})^{2}} \right) ^{7/2} \Upsilon(\tau_{sw})\ ,
\label{eq:soundwaves}
\end{eqnarray}
where $f_{\text{sw}}$ denotes the present peak frequency of the spectrum,
\begin{equation}
f_{\textrm{sw}} \ = \ 
1.9\times10^{-5}\frac{1}{\tilde{v}_W}\left(\frac{\beta}{H_{n}} \right) \left( \frac{T_{n}}{100\textrm{GeV}} \right) \left( \frac{g_{\ast}}{100}\right)^{1/6} \textrm{Hz} \,.
\label{fsw}
\end{equation}
 The parameter $\alpha$ measures the released vacuum energy relative to the radiation energy density,
\begin{equation}
	\alpha
	=
	\frac{
		\Delta \left[V_{\mathrm{eff}}-		
		\frac{T}{4}\frac{dV_{\mathrm{eff}}}{dT}
		\right]_{T=T_n}
	}{
		\rho_{\mathrm{rad}}(T_n)
	},
	\qquad
	\rho_{\mathrm{rad}}(T_n)
	=
	\frac{3}{4}w_{\mathrm{false}}(T_n),
\end{equation}
where $\Delta X\equiv X_{\mathrm{true}}-X_{\mathrm{false}}$. The enthalpy density in each phase is obtained by
\begin{equation}
	w_j(T)
	=
	\rho_j(T)+p_j(T)
	=
	-T\frac{d}{dT}
	V_{\mathrm{eff}}\left(\phi_j(T),s_j(T),T\right),
	\qquad
	j\in{\mathrm{true},\mathrm{false}}.
\end{equation}
The inverse duration parameter $\beta$ is defined as
\begin{eqnarray}
\frac{\beta}{H_n}=T\frac{d (S_3(T)/T)}{d T}|_{T=T_n}\; ,
\end{eqnarray}
where $H_n$ denotes the Hubble parameter at $T_n$. 
The $\kappa_{v}$ characterizes the fraction of the latent heat converted into the kinetic energy of the plasma fluid \cite{Espinosa:2010hh},
\bea
\kappa_v &\simeq & \frac{c_s^{11/5} \kappa_A \kappa_B}{(c_s^{11/5} - \tilde{v}_W^{11/5}) \kappa_B + \tilde{v}_W c_s^{6/5} \kappa_A} ~~({\rm for}~ \tilde{v}_W <c_s),\nonumber\\
\kappa_v &\simeq & \kappa_B + (\tilde{v}_W-c_s) \delta\kappa +\frac{(\tilde{v}_W-c_s)^3}{(\xi_J-c_s)^3}\left[\kappa_C-\kappa_B-(\xi_J-c_s)\delta\kappa\right] ~({\rm for}~ c_s<\tilde{v}_W<\xi_J),\nonumber\\
 \kappa_{v} &\simeq & \frac{(\xi_J- 1)^3 \xi_J^{5/2} \tilde{v}_W^{-5/2} \kappa_C \kappa_D }{[(\xi_J -1)^3 - (\tilde{v}_W - 1)^3] \xi_J^{5/2} \kappa_C + (\tilde{v}_W - 1)^3 \kappa_D}  ~~({\rm for}~ \xi_J<\tilde{v}_W )
\eea
with the sound velocity $c_s=\sqrt{1/3}$ and
\bea
&&\kappa_A \simeq \frac{ 6.9 \alpha \tilde{v}_W^{6/5}}
{1.36- 0.037 \sqrt{\alpha}+ \alpha},~~~
\kappa_B \simeq \frac{\alpha^{2/5}}
{0.017 + (0.997 + \alpha)^{2/5}},\nonumber\\
&&\kappa_C \simeq \frac{\sqrt{\alpha}}
{0.135 + \sqrt{0.98+\alpha}},~~~
\kappa_D \simeq \frac{\alpha}
{0.73 + 0.083\sqrt{\alpha} +\alpha}
\nonumber\\
&&\xi_J \simeq  \frac{\sqrt{\frac{2}{3}\alpha+\alpha^2}+\sqrt{1/3}}{1 + \alpha},~~~\delta \kappa\simeq -0.9\log\frac{\sqrt{\alpha}}{1+\sqrt{\alpha}}.
\eea
The finite duration of the sound-wave period leads to a suppression of the GW signal. We account for this effect through the factor \cite{Guo:2020grp} 
\beq
\Upsilon(\tau_{sw})=1-\frac{1}{\sqrt{1+2\tau_{sw}H_n}},
\eeq 
where $\tau_{sw}$ denotes the lifetime of the sound waves,
\beq
\tau_{sw}=\frac{\tilde{v}_W(8\pi)^{1/3}}{\beta \bar{U}_f}, ~~ \bar{U}^2_f=\frac{3}{4}\frac{\kappa_v\alpha}{1+\alpha}.
\eeq

We estimate the bubble wall velocity, $\tilde v_W$, in the local thermal equilibrium (LTE) approximation using the fitting formula of Ref.~\cite{Ai:2023see}. 
 All thermodynamic quantities are evaluated at the nucleation temperature $T_n$.
The enthalpy ratio is
\begin{equation}
	\Psi_n
	=
	\frac{w_{\mathrm{true}}(T_n)}
	{w_{\mathrm{false}}(T_n)}.
\end{equation}
In terms of $\alpha$ and $\Psi_n$, the low velocity asymptotic solution is
\begin{equation}
	\tilde v_{\mathrm{low}}
	=
	\left[
	\frac{3\alpha+\Psi_n-1}
	{2\left(2-3\Psi_n+\Psi_n^3\right)}
	\right]^{1/2},
\end{equation}
whereas the high velocity approximation is
\begin{equation}
	\tilde v_{\mathrm{high}}
	=
	\xi_J
	\left[
	1-a\frac{(1-\Psi_n)^b}{\alpha}
	\right].
\end{equation}
The two asymptotic solutions are interpolated using a generalized $p$-norm,
\begin{equation}
	\tilde v_W
	=
	\left(
	\tilde v_{\mathrm{low}}^{p}
	+
	\tilde v_{\mathrm{high}}^{p}
	\right)^{1/p},
\end{equation}
with the fitted coefficients
\begin{equation}
	a=0.2233,
	\qquad
	b=1.704,
	\qquad
	p=-3.433.
\end{equation}
Since $p<0$, this interpolation provides a smooth approximation to the minimum of the low and high velocity solutions. We apply the fit only to parameter points satisfying
\begin{equation}
	0<\tilde v_W<\xi_J<1,
\end{equation}
corresponding to stationary deflagration or hybrid solutions within the LTE treatment.

Assuming a Kolmogorov-type turbulent spectrum \cite{Kosowsky:2001xp}, the GW contribution from the MHD turbulence is given by \cite{Caprini:2009yp,Binetruy:2012ze}
\begin{eqnarray}
\Omega_{\textrm{turb}}h^{2} & \ = \ & 
3.35\times10^{-4}\left( \frac{H_{n}}{\beta}\right)\left(\frac{\kappa_{\textrm{turb}} \alpha}{1+\alpha} \right)^{3/2} \left( \frac{100}{g_{\ast}}\right)^{1/3} \tilde{v}_W \nonumber \\
  && \times \frac{(f/f_{\textrm{turb}})^{3}}
  {[1+(f/f_{\textrm{turb}})]^{11/3}(1+8\pi f/h_{n})},
\label{eq:mhd}
\end{eqnarray}
where $h_n$ is the red-shifted Hubble rate at the time of GW production,
\beq
h_n=1.65\times 10^{-5}\left(\frac{T_n}{100\textrm{GeV}}\right) \left(\frac{g_{\ast}}{100}\right)^{\frac{1}{6}}\textrm{Hz}.
\eeq
 The peak frequency $f_{\rm turb}$ is 
\begin{equation}
f_{\textrm{turb}} \ = \ 2.7\times10^{-5}\frac{1}{ \tilde{v}_W}\left(\frac{\beta}{H_{n}} \right) \left( \frac{T_{n}}{100\textrm{GeV}} \right) \left( \frac{g_{\ast}}{100}\right)^{1/6} \textrm{Hz} \,.
\end{equation}
$\kappa_{\rm turb}$ is expected to account for approximately $5\%-10\%$ of the energy fraction carried by the bulk fluid motion \cite{Hindmarsh:2015qta}.
 Here we adopt the representative choice $\kappa_{\text{turb}} = 0.1 \kappa_v$.

The forthcoming space-based interferometers, such as  LISA, can provide an
opportunity to test the GW signals associated with the EWPT. We assess the observability of the predicted spectrum by
comparing it with the detector sensitivity through the signal-to-noise ratio,
\begin{equation}
\mathrm{SNR}
=
\sqrt{
T\int_{f_{\min}}^{f_{\max}}
\mathrm{d}f\,
\left[
\frac{\Omega_{\rm GW}(f)}
{\Omega_{\rm LISA}(f)}
\right]^2
}.
\end{equation}
Here, $\Omega_{\rm LISA}(f)$ represents the LISA sensitivity curve~\cite{Caprini:2015zlo},
while $T$ denotes the observation time. We assume a four-year mission,
corresponding to $T\simeq1.26\times10^{8}~{\rm s}$.
For the six-link LISA configuration, we adopt
$\mathrm{SNR}\geq 10$ as the criterion for detectability.

\subsection{Results and discussions}
\begin{figure}[tb]
\begin{center}
 \epsfig{file=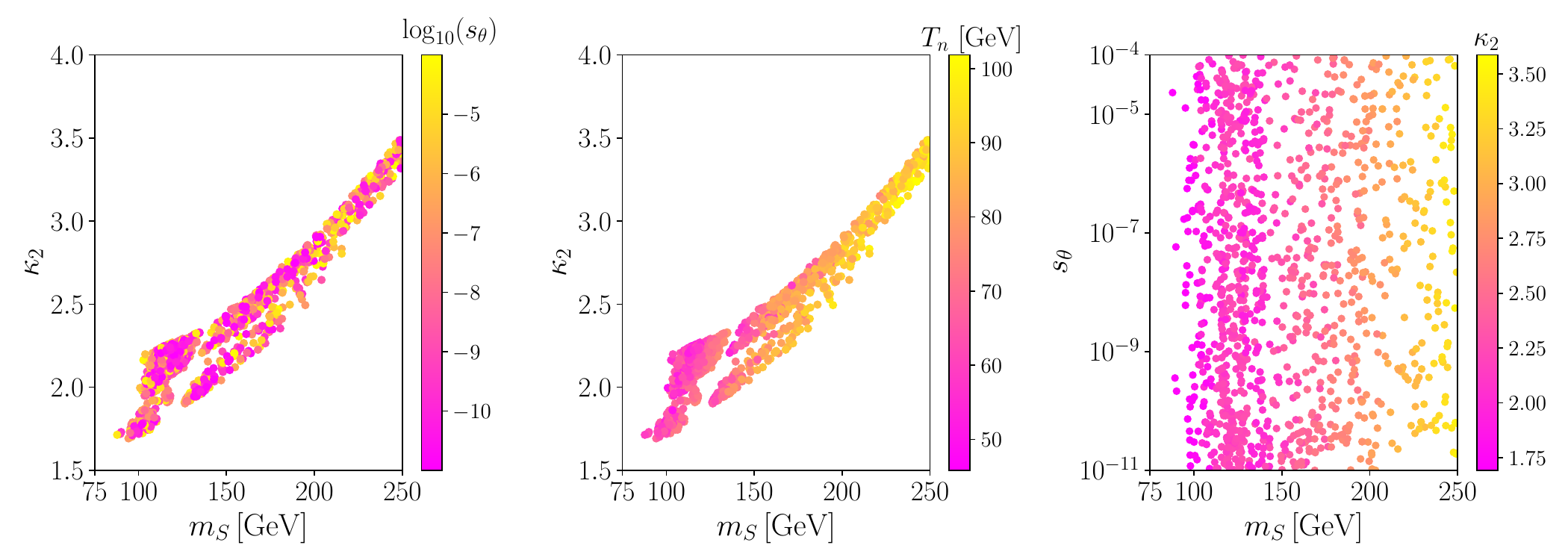,height=5.5cm}
 \end{center}
\vspace{-1.0cm} \caption{The parameter points realize an SFOEWPT while satisfying the relevant constraints.} \label{tn-pt}
\end{figure}

After applying the theoretical constraints, we identify the parameter points that realize an SFOEWPT in Fig. \ref{tn-pt}. 
In the high-temperature regime, thermal effects tend to drive $\langle \phi\rangle$ toward zero, 
thereby restoring the electroweak symmetry. In contrast,
 the singlet field generally acquires a non-vanishing  expectation value due to the presence of a thermal tadpole. 
Since the model does not possess $s \to -s$ symmetry, 
there is no symmetry requirement for $\langle s\rangle$ to vanish in the high-temperature limit.

\begin{figure}[tb]
\begin{center}
 \epsfig{file=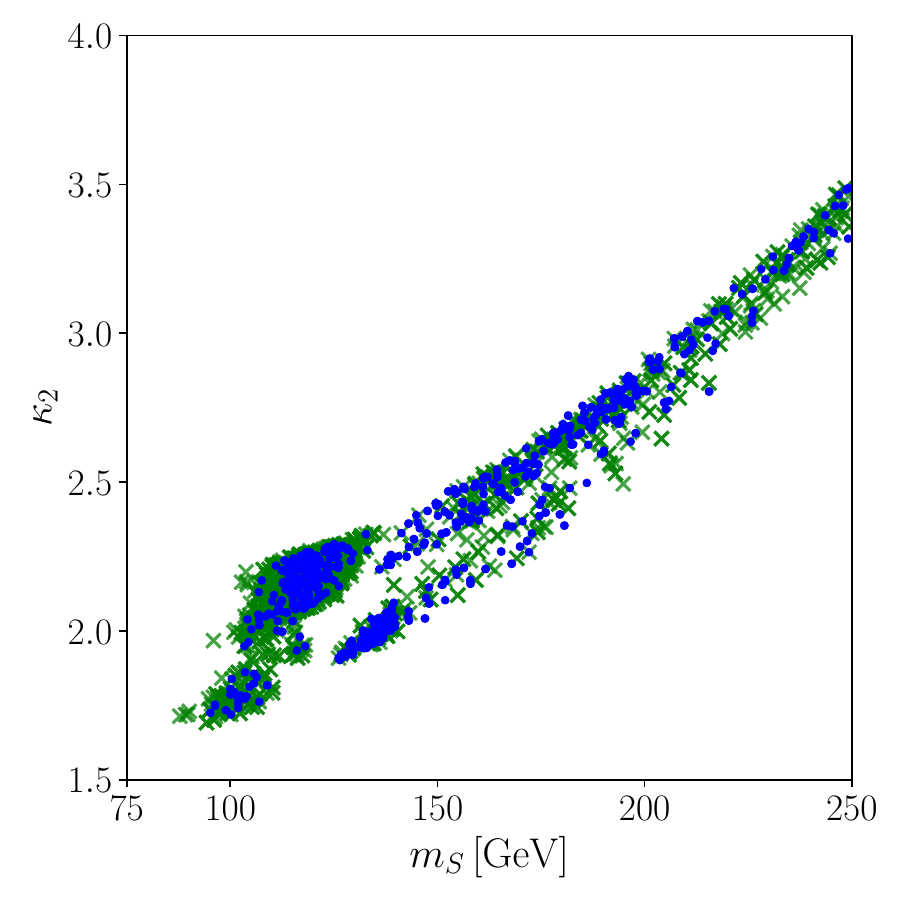,height=7.0cm}
 \end{center}
\vspace{-1.0cm} \caption{All the parameter points realize an SFOEWPT while satisfying the relevant constraints. The green and blue points denote SNR $<10$ and SNR $\geq 10$ for the detectability of the GWs at the LISA, respectively. } \label{fig:gw}
\end{figure}

Fig.~\ref{tn-pt} shows that the parameter space realizing an SFOEWPT is sensitive to $m_S$ and $\kappa_2$. 
The surviving points are concentrated in the ranges $80~\mathrm{GeV}<m_S<250~\mathrm{GeV}$ and $1.6<\kappa_2<3.5$. 
The left panel reveals a clear positive correlation between $m_S$ and $\kappa_2$,
 indicating that $\kappa_2$ tends to increase with $m_S$.
This correlation reflects the role of $\kappa_2$ in shaping the scalar potential and the thermal dynamics of the Higgs-singlet system. 
Only specific combinations of $m_S$ and $\kappa_2$ can realize an SFOEWPT.
Also, the nucleation temperature $T_n$ tends to increase with $m_S$ and $\kappa_2$, as shown in the middle panel.

In contrast, the SFOEWPT shows little dependence on the mixing angle $s_\theta$. 
The viable points span the entire $s_\theta$ range, 
with no significant correlation between $s_\theta$ and either $m_S$ or $\kappa_2$. 
This is particularly relevant for the long-lived scalar scenario, 
as a small $s_\theta$ can suppress the decay width of the singlet scalar while still allowing an SFOEWPT.

We further classify the parameter points in Fig.~\ref{tn-pt} according to their expected GW detectability at the LISA. As shown in Fig.~\ref{fig:gw}, the green points correspond to $\mathrm{SNR}<10$, while the blue points satisfy $\mathrm{SNR}\geq10$, which we take as the criterion for detectability at the LISA. The blue points extend over a broad range of $m_S$, indicating that part of the parameter space realizing an SFOEWPT can produce GW signals within the projected sensitivity of LISA.

\section{Searches for long-lived scalars at the LHC}
\label{sec:lhc}

The small Higgs-singlet mixing strongly suppresses the couplings of $S$ to the SM fermions and gauge bosons, 
whereas the $hSS$ interaction remains sizable. Consequently, above the SM-like Higgs decay threshold, 
the dominant production channel of $S$ at the LHC is
$gg\to h^*\to SS$, mediated by an off-shell SM-like Higgs boson.

We construct the phenomenological model using
\texttt{FeynRules}~\cite{Christensen:2008py,Alloul:2013bka} and
\texttt{NLOCT}~\cite{Degrande:2014vpa}, and export the resulting interactions in
the UFO format. The UFO model is then imported into
\texttt{MadGraph5aMC$@$NLO}~\cite{Alwall:2014hca,Frederix:2018nkq}
to generate the LO parton-level events.
For the process $gg\to h^*\to SS$, the production rate is evaluated with the
full one-loop matrix element, including the exact top-quark mass dependence.
 The renormalization and factorization scales are both set to
$\mu_R=\mu_F=m_{SS}$,
where $m_{SS}$ is the invariant mass of the scalar pair.
Our primary goal is to assess the sensitivity to long-lived scalar signatures rather than to obtain a precision prediction of the inclusive production rate. 
We therefore do not pursue a detailed treatment of higher-order QCD corrections,
 but instead approximate the dominant NLO effects by adopting a constant $K$-factor of $1.5$, 
following Refs.~\cite{Borowka:2016ypz,Baglio:2018lrj,Hespel:2014sla}.
The generated LHE event samples~\cite{Alwall:2006yp} are subsequently
processed with \texttt{Pythia8}~\cite{Sjostrand:2014zea,Bierlich:2022pfr} for parton showering and hadronization.
The scalar $S$ is decayed in \texttt{Pythia8} through all available channels,
with the corresponding branching fractions taken into account.

The decay location of a long-lived scalar depends on both its lifetime and boost. 
It may decay inside the main detector or travel beyond it and decay in a far detector, allowing it to be searched at either the main or far detectors.

\subsection{Signatures of displaced vertices + jets at  ATLAS}

The macroscopic lifetime of $S$ allows it to decay at a displaced location
within the tracking volume of the ATLAS or CMS detector. Its decay pattern
depends on the scalar mass: the dominant final states are $b\bar b$ at lower
masses and $W^+W^-$ or $ZZ$ once the weak-boson channels become important.
The hadronic decays of the gauge bosons further give rise to multiple jets.
The characteristic signature is therefore  displaced vertices (DVs) accompanied by
several jets.

This topology has been explored by ATLAS using the full Run-2 dataset,
corresponding to an integrated luminosity of $139~{\rm fb}^{-1}$
\cite{ATLAS:2023oti}. The original search targets long-lived electroweakinos
in R-parity-violating supersymmetry, which decay into multi-jet final states. No signal excess was observed,
and the analysis placed constraints on the electroweakino masses and
lifetimes.
The same search has subsequently been adapted to several other LLP
scenarios. Ref.~\cite{Cheung:2024qve}, following the ATLAS recasting
prescription~\cite{atlas_recast_instruction}, reinterpreted the analysis for
an axion-like particle with quark-flavor-violating couplings. The validation
performed in Ref.~\cite{Cheung:2024qve} covers both light- and heavy-flavor
jet final states. This recast framework has  been employed in
Refs.~\cite{Wang:2024ieo,Beltran:2025ilg,Wang:2025zhd}. 
A similar recasting of the same search was also performed in Ref.~ \cite{Heisig:2024xbh}.


A modified version of the ATLAS strategy was introduced in
Ref.~\cite{Beltran:2025ilg}. The main difference is a relaxation of the
jet-$p_T$ requirements, motivated by the earlier 8-TeV ATLAS DV search
\cite{ATLAS:2015oan}, while the DV selection and reconstruction efficiencies
are retained from the 13-TeV analysis~\cite{ATLAS:2023oti}. The motivation
for this modification is that the main background rejection is expected to
come from the DV reconstruction and vertex-level requirements. 

For the detector-level treatment, the recast procedure employs a simplified
toy-detector module implemented within \texttt{Pythia8}~\cite{Sjostrand:2014zea,Bierlich:2022pfr},
following the ATLAS recast prescription~\cite{atlas_recast_instruction}.
Truth-level jets are reconstructed with \texttt{FastJet}~\cite{Cacciari:2011ma}
using the anti-$k_t$ algorithm with $R=0.4$, with the same definition adopted
for both prompt and displaced jets. The detector response to the jet
transverse momenta is modeled according to Ref.~\cite{Allanach:2016pam},
including detector acceptance, momentum resolution, and $p_T$ smearing.

We apply both versions of the analysis to our scenario, and refer to the two implementations as the ``original''
and ``modified'' analyses, respectively. 
The event-level requirements of the two analyses are summarized in
Table~\ref{tab:jet_pt_requirement}. Both begin with a selection based on the
number of truth jets satisfying specified transverse-momentum thresholds.
For the original analysis, additional requirements are imposed on the number
of displaced jets. A displaced jet is identified by associating it with an
LLP decay according to the angular separation between the decay products and
the corresponding truth jet.

\begin{table}[t]
\begin{center}
\begin{tabular}{c|c}
            Original analysis           & Modified analysis                                                             \\ \hline
               $n^{137}_{\text{jet}}\geq 4$ or $n^{101}_{\text{jet}}\geq 5$    & $n^{90}_{\text{jet}}\geq 4$ or $n^{65}_{\text{jet}}\geq 5$                   \\
 or $n^{83}_{\text{jet}}\geq 6$ or $n^{55}_{\text{jet}}\geq 7$, & or $n^{55}_{\text{jet}}\geq 6$                 \\
           $n^{70}_{\text{displaced jet}}\geq 1$ or $n^{50}_{\text{displaced jet}}\geq 2$                                                           &  \\ \hline
\end{tabular}
\caption{Event-level acceptance requirements based on the truth-jet transverse
momenta. Here, $n^{137}_{\text{jet}}$, for example, denotes the number of
truth jets with $p_T\geq137$ GeV.}
\label{tab:jet_pt_requirement}
\end{center}
\end{table}

Events satisfying the jet requirements are subsequently subjected to the
DV-level selection. At least one LLP decay in the event must produce a
reconstructed DV satisfying all of the following conditions:
\begin{enumerate}
    \item \textbf{Fiducial volume:}
    \[
    4~{\rm mm}<R_{xy}<300~{\rm mm},
    \qquad
    |z|<300~{\rm mm},
    \]
    where $R_{xy}$ and $|z|$ denote the transverse and longitudinal
    distances of the DV from the interaction point (IP), respectively.

    \item \textbf{Track impact parameter:}
    at least one track associated with the DV must satisfy
    $    |d_0|>2~{\rm mm}$,
    where $d_0$ is the transverse impact parameter.

    \item \textbf{Decay-product requirements:}
    the DV must contain at least five massive decay products satisfying
    \begin{enumerate}
        \item a transverse decay length in the laboratory frame,
        $\beta_T\gamma c\tau$, larger than $520~{\rm mm}$, where $\beta_T$
        denotes the transverse velocity, $\gamma$ is the Lorentz factor,
        and $c\tau$ is the proper decay length;
        \item
        $  \frac{p_T}{|q|}>1~{\rm GeV}$,
              where $p_T$ and $q$ are the transverse momentum and electric charge
        of the decay product.
    \end{enumerate}

    \item \textbf{DV invariant mass:}
    $    m_{\rm DV}>10~{\rm GeV}$.
       The invariant mass is reconstructed from the selected decay products,
    treating each of them as a charged pion.
\end{enumerate}

The ATLAS-provided parameterized efficiencies are then applied to the events
that survive the above acceptance requirements \cite{atlas_recast_instruction}. These efficiencies account
for detector and reconstruction effects that are not explicitly simulated
in the recast. For the original analysis, both event-level and vertex-level
efficiencies are included, whereas only the vertex-level efficiencies are
used for the modified analysis. The event-level efficiency depends on the
scalar sum of the truth-jet transverse momenta and the transverse position
$R_{xy}$ of the most distant LLP decay. The vertex-level efficiency is
parameterized in terms of $R_{xy}$, $m_{\rm DV}$, and the multiplicity of
the LLP decay products. 

The dominant sources of background include the artificial merging of nearby
low-mass DVs, hadronic interactions with detector material, and accidental
associations of tracks with unrelated low-mass vertices. Nevertheless, the
selection requirements suppress these backgrounds to a very small level.
For the original analysis considered here, ATLAS estimates a
background yield of
$0.83^{+0.51}_{-0.53}$ events, while no events were observed in the data.

For the original analysis, we recast the ATLAS search using the full Run-2 dataset, 
corresponding to an integrated luminosity of $139~{\rm fb}^{-1}$. 
For the modified analysis, we project the sensitivity to the HL-LHC, assuming an integrated luminosity of $3~{\rm ab}^{-1}$.


\subsection{Searches with LHC far detectors}
Several dedicated far-detector experiments have been proposed to extend
the LHC sensitivity to LLP with macroscopic decay
lengths. By providing fiducial decay volumes at substantial distances from
the IP, these detectors can probe LLP decays occurring
outside the main ATLAS and CMS detectors. In this work, we consider the
following far-detector configurations: ANUBIS, CODEX-b, FACET, 
FASER2, MoEDAL-MAPP1 and MAPP2, and MATHUSLA. Their respective geometrical
configurations are implemented in the \textsf{Displaced Decay Counter (DDC)} \cite{Domingo:2023dew}, which
is used to evaluate the corresponding LLP decay acceptances.

Among these detectors,  FASER2 and FACET are located in the forward
region along the LHC beam direction. In particular, FASER2 is designed to detect LLPs produced in a very narrow cone around the beam axis. Its small transverse acceptance limits its sensitivity to the scalar $S$
in the present scenario, since the decay products of
$S$ are expected to be distributed predominantly in the transverse
direction. 

It is worth noting that the geometrical configurations of MATHUSLA and
ANUBIS have undergone significant updates. The earlier MATHUSLA design
featured a fiducial volume of
$100~\mathrm{m}\times100~\mathrm{m}\times25~\mathrm{m}$.
In the latest design, the fiducial volume is reduced to
$40~\mathrm{m}\times40~\mathrm{m}\times11~\mathrm{m}$.
The detector is located near the CMS IP, with horizontal
and vertical distances of approximately $70~\mathrm{m}$ and $81~\mathrm{m}$
from the CMS IP, respectively~\cite{MATHUSLA:2025zyt}.
The original ANUBIS-shaft configuration was designed to be installed in a
service shaft above the ATLAS IP, with a cylindrical
fiducial volume of height $56~\mathrm{m}$ and diameter $18~\mathrm{m}$.
The updated design instead exploits the space between the ATLAS detector
and the cavern ceiling. Its fiducial volume is described by an annular
cylindrical sector with a length of $53~\mathrm{m}$, inner and outer radii
of $11.3~\mathrm{m}$ and $19.3~\mathrm{m}$, respectively, and an azimuthal
coverage of approximately $20\%$~\cite{ANUBIS:2025sgg}.
Compared with the previous shaft-based configuration, ANUBIS-ceiling is
located closer to the ATLAS IP, resulting in an enhanced
geometric acceptance. We therefore adopt the updated MATHUSLA and ANUBIS
configurations in our analysis, which are implemented in the \textsf{DDC} \cite{Domingo:2023dew}.

\subsection{Results and discussions}

\begin{figure}[tb]
\begin{center}
 \epsfig{file=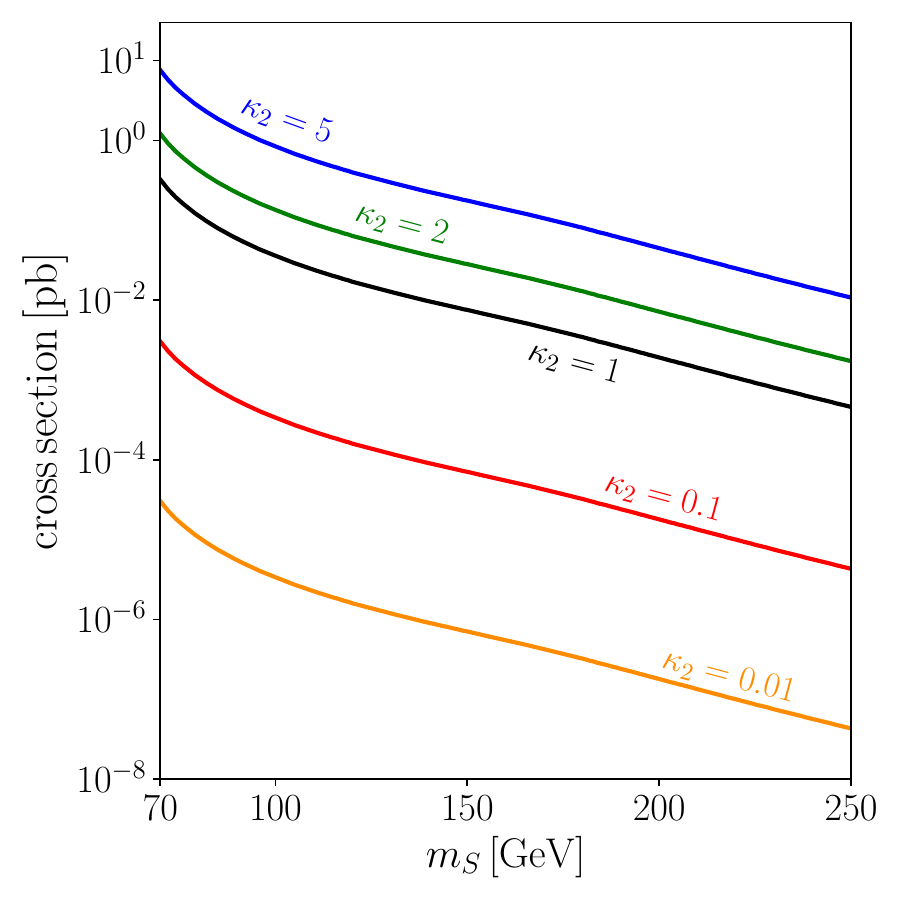,height=7.5cm}
 \end{center}
\vspace{-1.0cm} \caption{Cross sections 
for $gg\to h^*\to SS$  at the $\sqrt{s}=$ 14 TeV LHC for several values of $\kappa_2$.} \label{fig:cs}
\end{figure}

\begin{figure}[tb]
\begin{center}
 \epsfig{file=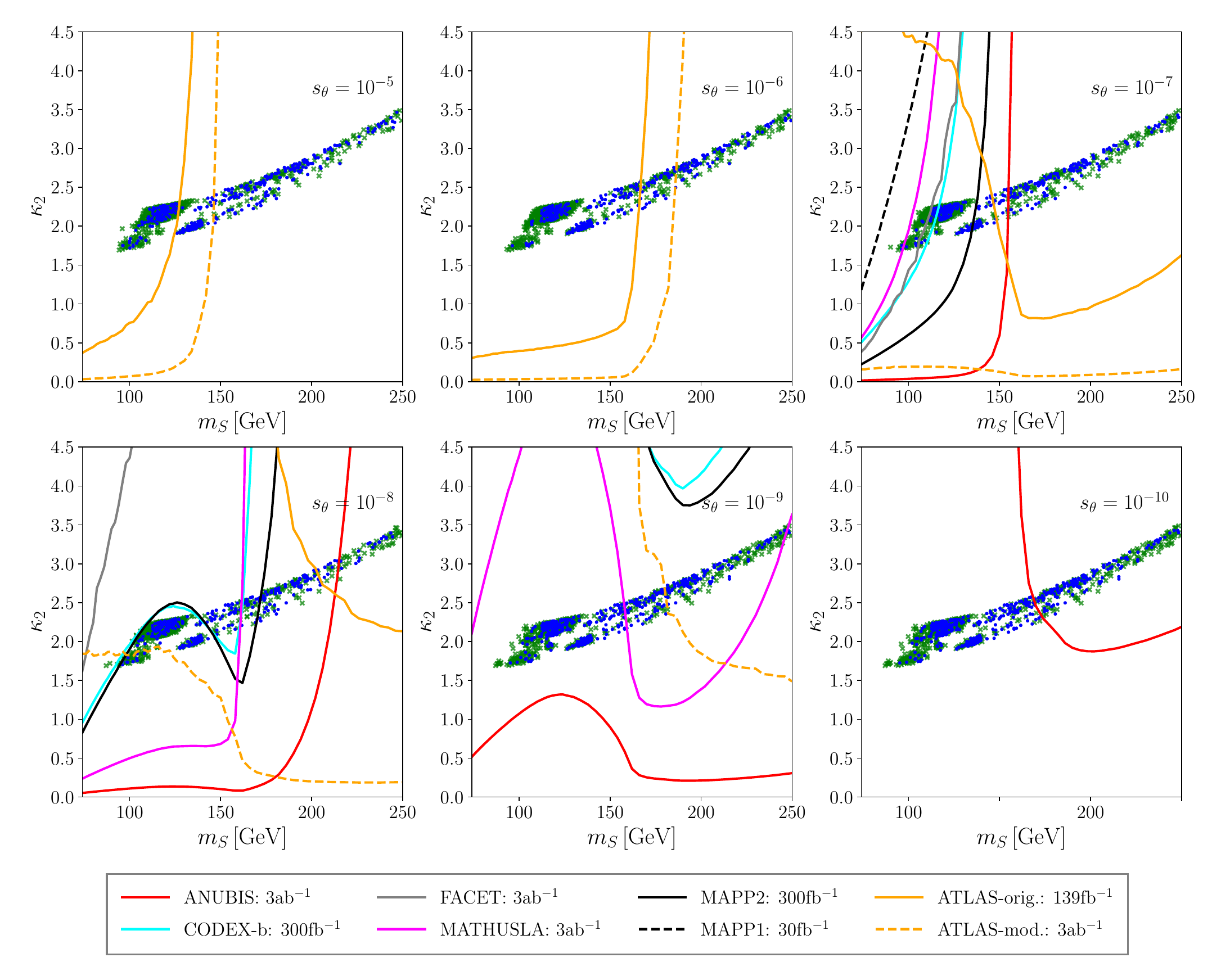,height=12.0cm}
 \end{center}
\vspace{-1.0cm} \caption{The 95\% C.L. sensitivity reach to $\kappa_2$ with respect to $m_S$. The scattered green and blue points have the same meanings as those shown in the Fig \ref{fig:gw}.} \label{fig:lhc}
\end{figure}

In Sec.~\ref{sec:decay}, we present the branching ratios of the different decay modes of $S$. Here we first discuss the cross section of the dominant $S$ production process at the LHC with $\sqrt{s}=14$ TeV, namely, $gg\to h^*\to SS$. Fig.~\ref{fig:cs} shows the corresponding production cross section. For a fixed $\kappa_2$, the cross section decreases as $m_S$ increases, mainly due to the reduced phase space available for producing the heavier scalar pair. In contrast, the production rate increases rapidly with $\kappa_2$, following $\sigma(gg\to h^*\to SS)\propto\kappa_2^2$, since $\kappa_2$ determines the $hSS$ coupling.


Fig.~\ref{fig:lhc} shows the 95\% C.L. sensitivity reach in the $(m_S,\kappa_2)$ plane for different values of the mixing angle, $s_\theta=10^{-5,-6,-7,-8,-9,-10}$. The colored curves represent the sensitivity projections of ATLAS and several dedicated far detectors, including ANUBIS, CODEX-b, FACET, MoEDAL-MAPP1 and MAPP2, and MATHUSLA. The 95\% C.L. sensitivity is defined by requiring approximately three signal events.
The scattered green and blue points can achieve an SFOEWPT, while for the blue points the GW signals within the projected sensitivity of LISA. 
As discussed above, the highly forward geometry and narrow angular acceptance of FASER2 strongly limit its sensitivity to the scalar S in the present scenario. As expected,  FASER2 therefore provide no sensitivity to the SFOEWPT region.

The production cross section through $gg\to h^\ast\to SS$ scales as $\kappa_2^2$. Therefore, for fixed $m_S$ and $s_\theta$, a larger $\kappa_2$ enhances the number of produced $S$ particles and can compensate for a reduced acceptance. 
On the other hand, the lifetime of $S$ is controlled by $s_\theta$ and $m_S$. Since its total decay width is proportional to $s_\theta^2$, and the proper decay length $c\tau_S$ increases rapidly as $s_\theta$ becomes smaller.
 At the same time, the decay width of $S$ increases with its mass. In particularly, in the high-mass regime, the opening of the $S\to W^+W^-$ and $S\to ZZ$ channels further enhances the total decay width, resulting in a more pronounced mass dependence of the lifetime.

For the relatively large mixing angles, $s_\theta=10^{-5}$ and $10^{-6}$, a sizable fraction of the produced $S$ particles decay within the
central ATLAS detector, whereas the probability for them to survive to
the locations of the far detectors is strongly suppressed. The original
ATLAS analysis therefore provides sensitivity mainly to the region in
which the lifetime of $S$ is sufficiently long to yield DVs, but not so long that most of the particles escape the ATLAS
fiducial volume. The future modified ATLAS analysis has a substantially
improved reach. In particular, lowering the transverse-momentum
threshold in the jet selection enhances the signal efficiency, while
the assumed integrated luminosity is increased from
$139~\mathrm{fb}^{-1}$ to $3~\mathrm{ab}^{-1}$. The modified analysis
therefore extends the sensitivity toward smaller values of $\kappa_2$
and a wider range of $m_S$ compared with the original analysis.
At the high-mass region, however, the increasing decay width of $S$ leads to a progressively shorter lifetime, particularly with the opening of the $WW$ and subsequently $ZZ$ decay channels. As a result, $S$ may decay too promptly to yield an efficient DV signature, causing the
ATLAS DV searches to lose sensitivity to the high-mass
SFOEWPT region.

As $s_\theta$ decreases to $10^{-7}$, the lifetime of $S$ becomes longer, allowing an increasingly large fraction of the particles to escape the central detector before decaying, particularly in the low-mass region. Consequently, the far detectors begin to provide substantial sensitivity to the low- and intermediate-mass regions that realize an SFOEWPT. Meanwhile, the original ATLAS analysis retains sensitivity mainly in the intermediate- and high-mass regions, where the increase in the decay width associated with larger $m_S$ can partially compensate for the width suppression induced by the smaller $s_\theta$. The future modified ATLAS analysis provides the broadest coverage among the LHC projections and can cover essentially the entire SFOEWPT region shown in this panel.

For $s_\theta=10^{-8}$, the longer lifetime of $S$ noticeably reduces the sensitivity of both the original and modified ATLAS analyses, resulting in reduced coverage of the SFOEWPT parameter space. Among the far detectors considered, ANUBIS and MATHUSLA provide sizeable coverage of the SFOEWPT region, whereas MAPP1 and FACET lose sensitivity. These differences arise from the probability for $S$ to reach and decay within each detector, as well as from their different geometries and integrated luminosities.

For $s_\theta=10^{-9}$, the complementarity between the main and far detectors becomes more pronounced. The original ATLAS analysis no longer covers the SFOEWPT region, while the future modified ATLAS analysis retains sensitivity mainly at high $m_S$. The larger decay width at higher scalar masses can shorten the $S$ lifetime, allowing a fraction of $S$ particles to decay inside ATLAS. In contrast, ANUBIS covers the entire SFOEWPT region shown in this panel, while MATHUSLA remains sensitive mainly at high $m_S$. The other dedicated far detectors do not reach the SFOEWPT region in this case.

As $s_\theta$ decreases to $10^{-10}$, the lifetime of $S$ becomes very long across most of the parameter space. Consequently, the decay probability inside the central ATLAS detector is strongly suppressed, and the ATLAS searches no longer cover the SFOEWPT region. Among the dedicated far detectors, only ANUBIS retains sensitivity, mainly to the high-mass portion of the SFOEWPT region. The other far detectors lose sensitivity because a large fraction of the $S$ particles can traverse their fiducial volumes without decaying.

The comparison among the different panels demonstrates a clear complementarity between ATLAS and the dedicated far detectors. For relatively large $s_\theta$, corresponding to a shorter decay length of $S$, the sensitivity is dominated by ATLAS. As $s_\theta$ decreases and $S$ becomes increasingly long-lived, the sensitivity progressively shifts toward detectors with larger baselines and dedicated decay volumes, particularly ANUBIS and MATHUSLA. Each detector is most sensitive within a characteristic lifetime range determined by its geometrical configuration. Consequently, the sensitivity curves exhibit characteristic bends, which can also give rise to non-monotonic behavior in the $(\kappa_2,m_S)$ plane.

The results for $s_\theta=10^{-11}$ and $10^{-4}$ are not shown. Our detailed simulations and analysis indicate that, for $s_\theta=10^{-11}$, $S$ is too long-lived to decay efficiently within the fiducial volumes of ATLAS and the dedicated far detectors, leaving the SFOEWPT region essentially uncovered. In contrast, for $s_\theta=10^{-4}$, $S$ decays too promptly for DV searches to be effective, and the SFOEWPT region is likewise not covered.


\section{Conclusion}
\label{sec:conclu}
We have studied an SFOEWPT triggered by a long-lived $S$ in the singlet scalar extension of SM, 
together with its collider phenomenology at the LHC. The finite-temperature analysis shows that the SFOEWPT region is sensitive to $m_S$ and $\kappa_2$,
 with very little dependence on the mixing angle $\theta$.

At the LHC, we consider the pair-production process $gg\to h^*\to SS$ with an off-shell $h$. If $S$ decays within the inner detector, the resulting signature is DVs accompanied by multiple jet. If $S$ escapes the main detector, its subsequent decay can instead be detected by dedicated far detectors, including ANUBIS, CODEX-b, FACET, MoEDAL-MAPP1 and MAPP2, and MATHUSLA. The combined sensitivity of the main and far detectors, particularly ANUBIS and MATHUSLA, allows a sizable fraction of the SFOEWPT parameter space with $10^{-10}\leq\sin\theta\leq 10^{-5}$ and $m_S\leq 250~\mathrm{GeV}$ to be probed. In contrast, scenarios with $\sin\theta\leq 10^{-11}$ remain beyond the reach of the detector configurations considered here.

For part of the viable parameter space, the predicted GW spectra from the SFOEWPT fall within the projected sensitivity of LISA, including some parameter points beyond the reach of the LHC searches. This highlights the complementarity between GW observations and collider searches in probing the long-lived scalar scenario.

\acknowledgments
We are grateful to Zeren Simon Wang for helpful discussions. This work was supported by the Projects No. ZR2024MA001 and
No. ZR2023MA038 supported by Shandong Provincial 
Natural Science Foundation and by the National Natural Science Foundation
of China under grants No.11975013.

\appendix
\section{Field dependent mass $\hm_i(\phi,s)$ and thermal Debye masses $\bar{M}_i^2(\phi,s,T)$ }
In terms of the classical background fields $(\phi,s)$, the field dependent mass $\hm_i$ are given by
\bea
\hm_h^2\left( \phi,s\right) &=& \frac{1}{2}\left(m^2_{11} + m^2_{22} +\sqrt{(m^2_{11}-m^2_{22})^2 + 4(m^2_{12})^2}\right) ,\nonumber\\
\hm_S^2\left( \phi,s\right) &=&  \frac{1}{2}\left(m^2_{11} + m^2_{22} -\sqrt{(m^2_{11}-m^2_{22})^2 + 4(m^2_{12})^2}\right) ,\nonumber\\
\hm_G^2\left( \phi,s\right) &=& \hm_{G^\pm}^2\left( \phi,~s\right) =-\mu_h^2 + \frac{\lambda_1}{2} \phi^2 + \frac{\kappa_1}{2} s+ \frac{\kappa_2}{2} s^2  ,\nonumber\\
\hm_{W^\pm}^2 \left( \phi,s\right)&=& \frac{1}{4} g^2 \phi^2,\nonumber\\
\hm_{Z}^2\left( \phi,s\right) &=& \frac{1}{4} \left(g^2+g'^2\right)\phi^2,\nonumber\\
\hm_{t}^2\left( \phi,s\right) &=&  \frac{m_t^2}{v^2}\phi^2,
\eea
with
\bea
m^2_{11}&=&-\mu_h^2 + \frac{3\lambda_1}{2} \phi^2 + \frac{\kappa_1}{2} s+ \frac{\kappa_2}{2} s^2,\nonumber\\
m^2_{22}&=&\mu_2 + \frac{3\lambda_2}{2} s^2 + 2 \mu_3 s + \frac{\kappa_2}{2} \phi^2,\nonumber\\
m^2_{12}&=&\frac{\kappa_1}{2} \phi + \kappa_2 \phi s.
\eea

The thermal Debye masses $\bar{M}_i^2$ are given by, 
\bea
\label{eq:thermalmass}
\bar{M}_{h}^{2}\left( \phi,s,T\right) &=&\frac{1}{2}\left(m^2_{11} + m^2_{22} + \Pi_{\phi}+ \Pi_{s} +\sqrt{(m^2_{11}+ \Pi_{\phi}-m^2_{22}- \Pi_{s})^2 + 4(m^2_{12})^2}\right),\nonumber\\
\bar{M}_{S}^{2}\left( \phi,s,T\right) &=&\frac{1}{2}\left(m^2_{11} + m^2_{22} + \Pi_{\phi}+ \Pi_{s} -\sqrt{(m^2_{11}+ \Pi_{\phi}-m^2_{22}- \Pi_{s})^2 + 4(m^2_{12})^2}\right),\nonumber\\
\bar{M}_{G}^{2}\left( \phi,s,T\right) &=&-\mu_h^2 + \frac{\lambda_1}{2} \phi^2 + \frac{\kappa_1}{2} s+ \frac{\kappa_2}{2} s^2 + \Pi_{G},\nonumber\\
\bar{M}_{G^\pm}^{2}\left( \phi,s,T\right) &=&\bar{M}_{G}^{2}\left( \phi,s,T\right),
\eea
where $\Pi_i$ denotes the thermal mass terms of the field $i$, 
\begin{align}
\Pi_{\phi} &= \left[{9g^2\over 2} + {3g'^2\over 2} + {6y_t^2 } + 6\lambda_{1} + \kappa_{2} \right] {T^2 \over 24},\nonumber\\
\Pi_{s} &= \left[3\lambda_{2} +4\kappa_2 \right] {T^2 \over 24},\nonumber\\
\Pi_{G} &= \Pi_{\phi}.
\end{align}

The Debye masses of longitudinal gauge bosons are
\begin{align}
\bar{M}_{W^{\pm}_L}^2 \left(  \phi,s,T\right)&= {1 \over 4} g^2  \phi^2 + {11\over 6} g^2 T^2, \nonumber\\
\bar{M}_{Z_L,\gamma_L}^{2}\left( \phi,s,T\right) &=\frac{1}{8}\left(g^2+g'^2\right)\left( \phi^2+\frac{22}{3}T^2\right)
\pm \frac{1}{2}\Delta, 
\end{align}
with $\Delta=\sqrt{\left[\frac{1}{4}\left(g^2-g'^2\right)\left(\phi^2+\frac{22}{3}T^2\right)\right]^2+\frac{1}{4}g^4 g'^4  \phi^4}$.



\vspace{1cm}

\bibliographystyle{JHEP}
\bibliography{refs}

\end{document}